\documentclass[twocolumn,reprint,superscriptaddress,prl]{revtex4-1}  

 \usepackage{amssymb}
 \usepackage{dsfont}
 \usepackage{mathdots}
 \usepackage{amsmath}
\usepackage{graphicx}
\usepackage{dcolumn}
\usepackage{bm}
\usepackage{color}
\usepackage{soul}

\begin{document}

\title{ Interferometry with Photon-Subtracted Thermal Light} 

\author{Seyed Mohammad Hashemi Rafsanjani}
\email{shashem2@ur.rochester.edu}
\author{Mohammad Mirhosseini}
\author{Omar S. Maga\~{n}a-Loaiza}
\affiliation{Institute of Optics, University of Rochester, Rochester, New York 14627}

\author{Bryan T. Gard}
\affiliation{Hearne Institute for Theoretical Physics and Department of Physics and Astronomy, Louisiana State University, Baton Rouge, LA 70803}

\author{Richard Birrittella}
\affiliation{Department of Physics and Astronomy, Lehman College, The City University of New York, Bronx, New York 10468}

\author{B. E. Koltenbah}
\author{C. G. Parazzoli}
\author{Barbara A. Capron}
\affiliation{Boeing Research \& Technology, Seattle, WA 98124}

\author{Christopher C. Gerry}
\affiliation{Department of Physics and Astronomy, Lehman College, The City University of New York, Bronx, New York 10468}

\author{Jonathan P. Dowling}
\affiliation{Hearne Institute for Theoretical Physics and Department of Physics and Astronomy, Louisiana State University, Baton Rouge, LA 70803}

\author{Robert W. Boyd}
\affiliation{Institute of Optics, University of Rochester, Rochester, New York 14627}
\affiliation{Department of Physics, University of Ottawa, Ottawa, ON, K1N6N5, Canada}

\date{\today} 

\begin{abstract} 
We propose and implement a quantum procedure for enhancing the sensitivity with which one can
determine the phase shift experienced by a weak light beam possessing thermal statistics in passing
through an interferometer. Our procedure entails subtracting exactly one (which can be generalized
to m) photons from the light field exiting an interferometer containing a phase-shifting element in one of
its arms. As a consequence of the process of photon subtraction, and somewhat surprisingly, the
mean photon number and signal-to-noise ratio of the resulting light field are thereby increased, leading to
enhanced interferometry. This method can be used to increase measurement sensitivity in a variety
of practical applications, including that of forming the image of an object illuminated only by weak
thermal light.
\end{abstract}



\pacs{..........}

\maketitle

 \newcommand{\dg}[1]{#1^{\dagger}}
 \newcommand{\reci}[1]{\frac{1}{#1}}
 \newcommand{\ket}[1]{|#1\rangle}
 \newcommand{\nim}{\frac{1}{2}}
 \newcommand{\om}{\omega}
 \newcommand{\te}{\theta}
 \newcommand{\la}{\lambda}
 \newcommand{\nn}{\nonumber}                      
 \newcommand{\bra}[1]{\langle#1\vert}                 
 \newcommand{\ipr}[2]{\left\langle#1|#2\right\rangle}
  \newcommand{\up}{\uparrow}
  \newcommand{\down}{\downarrow}
  \newcommand{\dn}{\downarrow}         
  \newcommand{\hilight}[1]{\colorbox{yellow}{#1}}
  
\newcommand{\nbar}{\bar{n}}  
\newcommand{\ignore}[2]{\hspace{0in}#2}

Interferometry is the technique of choice for many of the most sensitive physical measurements to date \cite{PhysRevLett.116.061102}. It underlies many monumental discoveries in physics, such as Young's double slit experiment, the Michelson-Morley experiment that established the special theory of relativity, and recently and spectacularly in gravitational wave detection \cite{PhysRevLett.116.061102}. Apart from these foundational contributions, interferometry also plays an important role in many applications in optical metrology and imaging of astronomical objects, e.g. in stellar Michelson interferometry \cite{AAMichelson:1921wl,Monnier:2003kb}.
These applications often deal with sources of light that possess thermal statistics. The fluctuations in the number of photons of a thermal state is given by $\sqrt{\bar{n}(\bar{n}+1)}$ where $\bar{n}$ is the average number of photons contained in the field. For a dim source of thermal light, the magnitude of these fluctuations becomes comparable or even larger than $\bar{n}$,  overwhelming interferometric signals which are obtained by averaging the number of photons.

Ever since the advent of modern quantum optics, strategies have been developed to harness the quantum nature of light in order to enhance the accuracy of interferometric measurements \cite{Caves:1981ij,Grangier:1987ir, Giovannetti:810205,Giovannetti:2006cr,Gerry:2010dy,Giovannetti:2011fv}. These proposals use exotic quantum states of light, e.g. squeezed states or entangled states, to probe the physical process of interest \cite{Caves:1981ij, DAngelo:2001dq,Campos:2003ka,Mitchell:2004eg,Walther:2004ih,Nagata:2007jo,Kacprowicz:2010cw,Usuga:2010fw}. Unfortunately such an arrangement is infeasible when the object of interest is a remote source of light that possesses thermal statistics. Thus, it remains highly desirable to establish a protocol for increasing the sensitivity of interferometry by distilling the statistical information already contained in the collected light, rather than changing the nature of illumination.

Here we describe a means to enhance the phase sensitivity of interferometry based on the use of photon-subtracted thermal states, which are quantum states obtained by removing a fixed number of photons from a light field that possesses thermal statistics \cite{Parigi:2007fv}. Photon-subtracted states have recently attracted interest because of their applications in quantum communication, quantum computation, and quantum metrology \cite{Wenger:2004cw, Ourjoumtsev:2006im, NeergaardNielsen:2006hl, Ourjoumtsev:2009jh, Takahashi:2010kw,Parigi:2007fv, Rosenblum:2016jr, Carranza:2012fk,Birrittella:2014eg}. 
In contrast with the conventional approach in utilizing quantum states of light in metrology, we propose to implement the photon-subtraction immediately before detection.
We demonstrate that such a subtraction scheme leads to an enhancement in both the magnitude of the signal and the signal-to-noise ratio. 

The essence of how photon subtraction enhances both signal strength and the signal-to-noise ratio (SNR) can be captured by a simple model. An interferometer can be described by a phase-dependent unitary transformation that connects the 
annihilation operators of the input ports  ($\hat{a},\hat{b}$) to those of the output ports ($\hat{c},\hat{d}$). We assume a symmetric interferometer, that is, the beam splitters are 50\% transmitting and 50\% reflecting. 
We also assume that a controllable phase difference $\varphi$ is introduced between the two paths. The field operators of the output ports  are then related to those of the input ports through the following transformation:
\begin{align}\nn
\hat{c} &=\frac{1}{2} \left[(e^{i\varphi}-1)\hat{b}  + i(e^{i\varphi}+1)\hat{a} \right]\\
\hat{d} &=\frac{1}{2} \left[i(e^{i\varphi}+1)\hat{b}  + (1-e^{i\varphi})\hat{a} \right].
\end{align}

We assume that input port $\hat{a}$ is fed by thermal light and that port $\hat{b}$ is fed by the vacuum state. In this case the fields at both  output ports  possess thermal statistics, and their means and standard deviations are \cite{Gerry:2004tk}
\begin{align}\nn \label{equation_mean}
\bar{n}_c&= {\rm Tr}[\hat{c}^{\dagger}\hat{c} \hat{\rho}_0] =\bar{n}\cos^2 \frac{\varphi}{2},~~~~~
&\Delta n_c= \sqrt{\bar{n}_c^2+\bar{n}_c},\\ 
\bar{n}_d &= {\rm Tr}[\hat{d}^{\dagger}\hat{d} \hat{\rho}_0] =
\bar{n}\sin^2 \frac{\varphi}{2},~~~~~
&\Delta n_d= \sqrt{\bar{n}_d^2+\bar{n}_d}.
\end{align}
Here $\nbar = \text{Tr}[\hat{a}^\dagger \hat{a} \hat{\rho}_{th}]$ is the average occupation number in  input port $\hat{a}$. Note that although the initial density matrix describes a separable state of the form  $\hat{\rho}_0 = \hat{\rho}_{\rm{th}}^{(a)}\otimes\hat{\rho}_{\rm{vac}}^{(b)}$, the output cannot be written as a direct product of the two reduced density matrices of the two output ports. However the reduced density matrix for either of the output ports is itself that of a thermal state  \cite{Gerry:2004tk}. 

Our strategy for enhancing the measurement sensitivity is to suppress the vacuum contribution of the thermal light, which we do by the following means. The density matrix of  our initial state is given explicitly by
\begin{align}
\hat{\rho}_0 = \left(\sum_n  \frac{\nbar^n}{(1+\nbar)^{n+1}}\ket{n}_a\bra{n}_a \right) \otimes \ket{0}_b\bra{0}_b,
\end{align}
which describes a thermal state with an average occupation number of $\bar{n}$ in port $\hat{a}$ and the vacuum state in port $\hat{b}$. Here $\ket{0}_b$ is the vacuum state in port $\hat{b}$ and $\ket{n}_a$ is the Fock state of $n$ photons in port $\hat{a}$. Note that the most probable photon occupation number in a thermal state is the vacuum state $n=0$. However the vacuum does not produce any detection event. Therefore, if one can suppress the vacuum contribution to the thermal light field, the remaining events may be expected to provide  enhanced sensitivity. One means for suppressing the vacuum contribution is to make use of photon subtraction, which, somewhat counter intuitively, can increase the average number of photons in the resulting light field \cite{Parigi:2007fv,Zavatta:2009cn}. Subtracting a single photon from port $\hat{c}$ can be described by the following operation: 
\begin{align}
\hat{\rho}_0\rightarrow \hat{\rho}_1 = \frac{\hat{c}\hat{\rho}_0 \hat{c}^\dagger}{\text{Tr}[\hat{c}\hat{\rho}_0 \hat{c}^\dagger]} .
\end{align}
where the denominator of the last expression ensures normalization. 
After implementing the subtraction, the resulting reduced density matrix for port $\hat{c}$ reads (see supplementary material) 
\begin{align}
\hat{\rho}_{1c} = \sum_n  \frac{(n+1)\nbar_c^{n+1}}{(1+\nbar_c)^{n+2}}  \ket{n}_c ~_c\bra{n}.
\end{align}
One can readily confirm that for this photon-subtracted state the average photon number and its variance are given by 
\begin{align}\nn
&{\rm Tr}[\hat{c}^{\dagger}\hat{c} \hat{\rho}_1] = 2\bar{n}\cos^2 \frac{\varphi}{2}=2\nbar_c\\
&({\rm Tr}[(\hat{c}^{\dagger}\hat{c})^2 \hat{\rho}_1]- {\rm Tr}[\hat{c}^{\dagger}\hat{c} \hat{\rho}_1]^2)^{1/2} = \sqrt{2} \Delta n_c
\end{align}
Thus, conditioned on the subtraction of a single photon from port $\hat{c}$, the mean photon number in this port is doubled and the signal-to-noise ratio,  defined as the ratio of the mean photon number to its  standard deviation, is enhanced by a factor of $\sqrt{2}$. Interestingly, one can readily confirm that the same enhancement occurs for photons in port $\hat{d}$ after conditioning on subtraction of a single photon from port $\hat{c}$. Furthermore, one can show that removing a larger number of  photons from the input thermal state leads to an even more pronounced increase in the mean and SNR.

The simple model just described can be extended to include the possibility of loss and to cast the results in terms of the number of detection events. We then find, for example, that the average and the standard deviation of number of detection events associated with port $c$ can be expressed as \cite{[{An extensive theoretical analysis of our experiment is to be published separately: }] Parazzoli:2016ve}:
\begin{align}\nn
\bar{N}_c &= \bar{n}T \eta_2 \cos^2\frac{\varphi}{2}\left(\delta_{0m}+ \frac{(1-\delta_{0m})(m+1)}{ 1+\bar{n}(1-T)\eta_1\cos^2\frac{\varphi}{2}}\right)\\ 
\Delta N_c &= \bar{N}_c/\sqrt{\frac{(1+m)\bar{n}T \eta_2 \cos^2\frac{\varphi}{2}}{1+\bar{n}(T\eta_2+(1-T)\eta_1)\cos^2\frac{\varphi}{2}}}
\end{align}
Here $\eta_i$ is the detection efficiency at detector $i$, $T$ denotes the transmission of the beam splitter used for subtraction, $m$ is the number of subtracted photons, and $\delta_{0m}$ is the Kronecker delta function. 
Note that although reduced in magnitude, the enhancements in the mean and the SNR remain significant even in presence of realistic loss and limited detection efficiency.  
 
 \begin{figure}[tbp]
 \includegraphics[width=\columnwidth]{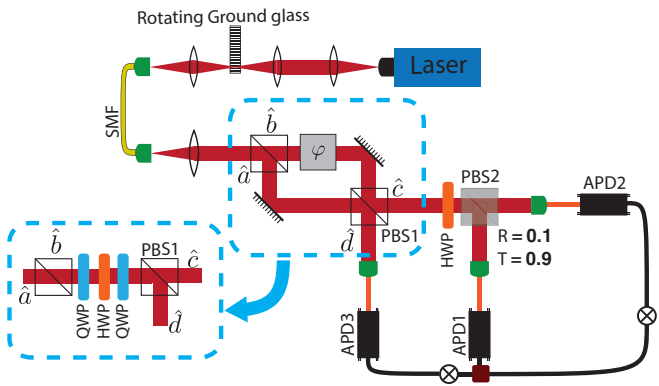} 
\caption{ Schematic of the experimental setup used to observe increased measurement sensitivity through photon subtraction. The output beam from a narrow-bandwidth, cw laser is focused onto a rotating ground-glass plate and is then coupled into a single-mode optical fiber (SMF). The single-transverse-mode, thermal light exiting the SMF is then sent to one input port ($\hat a$) of a Mach-Zehnder interferometer (MZI). Light from one output port ($\hat c$) is then sent to a combination of a half-wave plate (HWP) and polarizing beam splitter (PBS2) to perform the process of  photon subtraction. Detector APD2 counts the number of photons in a time window of fixed length, conditioned on a detection event in APD1. Similarly, light from the other output port ($\hat d$) is sent to detector APD3.  In our implementation (see inset), we use a common-path MZI to increase the stability. In this case a rotatable HWP is used to control the phase difference between two orthogonal polarization states of the light beam.}
\label{experimental_setup1}
\end{figure}
 \begin{figure}[tbp]
\includegraphics[width=0.8\columnwidth]{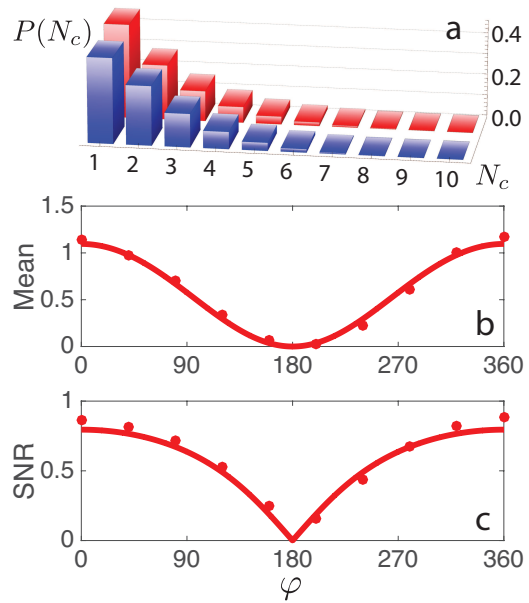}
\vspace{-0.3cm}
\caption{ (a) Histograms showing the photon number distribution in a single-mode thermal light field. Dark blue bars show laboratory results and red bars show theoretical predictions. (b) Mean photon number and (c) signal-to-noise ratio measured at the output of the interferometer by APD2 as a function of the phase difference between the two arms of the interferometer. The dots represent experimental results and the solid line  the theoretical prediction. Here the average photon number before the interferometer is $\bar{n} = 4.1$, the transmission of the subtracting PBS is $T=0.9$, the detection efficiency of APD2 is $\eta_2\approx 0.3$, and $\bar{N}_c = 1.1$}
\label{photon_statistics1}
\end{figure}

We next describe the experiment we performed to confirm the enhancement in measurement sensitivity. A schematic representation of our setup is shown in Fig.\,\ref{experimental_setup1}. We use a narrow-band external cavity diode laser operating at a wavelength of 780\,nm. The cw beam from the laser is focused onto a rotating ground-glass plate to produce pseudo-thermal light.  The beam is then coupled into a single-mode optical fiber to extract a single transverse mode of pseudothermal light. The light is then sent to a common-path Mach-Zehnder interferometer (MZI).  In this interferometer (see supplementary materials), the two arms correspond to the different phases acquired by two orthogonal polarization states of light.  We control these phases by means of a rotatable HWP placed within the interferometer. Without photon subtraction, our measurement sensitivity is limited by the standard fluctuations of thermal light.  We perform photon subtraction by diverting a small portion (10\%) of one of the outputs to detector APD1.  The fraction of the light sent to this detector is controlled by a HWP and polarizing beam splitter (PBS2).  A ``click" at this detector indicates that one photon has been removed from the beam. Similarly, two clicks within our integration time of $\simeq 1 \mu$s  indicates that two photons have been removed, etc.  We then count the number of detection events for detectors APD2 and APD3 conditioned on the number of detection events measured by APD1.

We perform photon counting through use of avalanche photodiodes (APDs) operating in the Geiger mode. The coherence time of our laser is approximately 1 $\mu s$,  and to ensure that we perform measurements on a single-temporal-mode field we use an  integration times equal to the coherence time.   The deadtime of our APD detectors is approximately 50 ns. To minimize errors associated with the arrival of a second photon within the deadtime following  a specific detection event, we adjust our laser intensity so that only a small number ($\lesssim 4$) of photons arrive in any one integration time. We also use statistical methods to correct our raw data for the rare occurrence of multiple photons arriving within one detector deadtime; see supplementary material for details.  Our use of a long-coherence-time light source allows us to time-bin the output of a standard APD to perform photon counting, thus circumventing the need to use photon-number-resolving detectors \cite{Zhai:2013in}. 

 In Fig.\ \ref{photon_statistics1} we show our experimental results for a case in which we do not implement photon subtraction and hence the light field possesses pure thermal statistics. We first (see part a) set the HWP in the MZI such that port $\hat{d}$ becomes dark and all the photons are directed toward port $\hat{c}$. We observe a negative exponential distribution (the signature of a thermal source) whose average occupation number is $\nbar T \eta_2  \sim 1.1$. Next we change the induced phase by rotating the HWP inside the MZI. For each value of the phase we find the average number of arriving photons (shown in part b) and its standard deviation. The signal-to-noise ratio (SNR) is then calculated  by dividing the average by the standard deviation (see part c).

 \begin{figure}[htbp]
 \includegraphics[width=0.8\columnwidth]{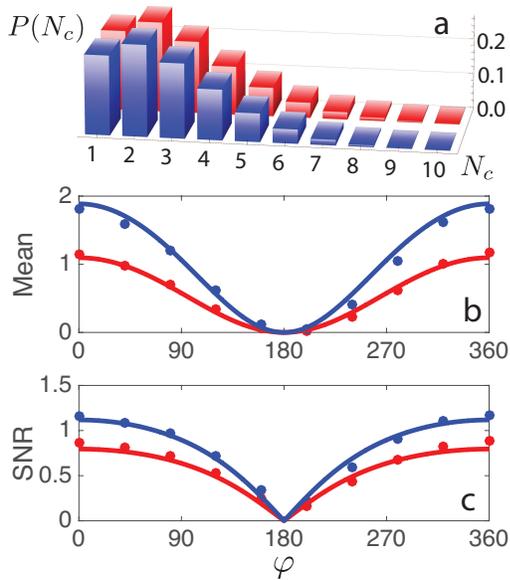}
 \vspace{-0.3cm}
\caption{(a) Histograms showing the photon number distribution for a single mode of one-photon-subtracted thermal light. Dark blue bars show laboratory results and red bars show theoretical predictions.  (b) Mean photon number and (c) signal-to-noise ratio measured at the output of the interferometer by APD2 as a function of the phase difference. The parameters are the same as Fig.\,\ref{photon_statistics1} and the detection efficiency of APD1 is $\eta_1\approx 0.33$. The dots represent experimental results and the solid line  the theoretical prediction. Also shown are the results for thermal light without subtraction (red line) for comparison.
}
\label{fig2}
\end{figure}


We next study the effect of photon subtraction on the sensitivity of the interferometer. In Fig.\ \ref{fig2}a,  we show a histogram of the photon number distribution measured at APD2 conditioned on the detection of a photon in the APD1 for $\varphi =0$. This statistical distribution is clearly different from the negative exponential distribution that we observe in Fig.\ 2a for thermal light. In parts b and c of the figure we show the mean photon number and the SNR plotted as functions of $\varphi$. For comparison, we also show the results for light with pure thermal statistics. We see that photon subtraction leads to an increase of the average photon number for $\varphi = 0$ from $1.1$ to $1.8$ and an increase in the SNR From $0.86$ to $1.15$.  The results are in very good agreement with the predictions of Eq.\,(7).  The increase in mean number is smaller than a factor of 2 and the increase in SNR is less than a factor of $\sqrt{2}$ as a consequence of loss. Nevertheless, even in presence of the loss we observe considerable enhancement in both signal and SNR. 

 \begin{figure}[htbp]
\includegraphics[width=0.8\columnwidth]{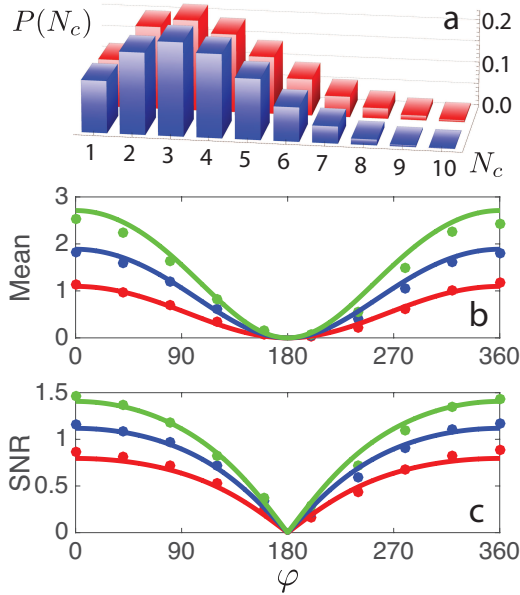} \vspace{-0.3cm}
\caption{Same as Fig.\,\ref{fig2}, except for two-photon-subtracted thermal light. Also shown are the results for thermal light without subtraction (red line) and a one-photon-subtracted thermal light (blue line) for comparison.
}
\label{fig3}
\end{figure}

Subtracting more than one photon leads to an even more pronounced increase in the mean photon number and SNR, as indicated by Eq.\,(7). We show results for the case of two-photon-subtracted thermal states in Fig.\,\ref{fig3}. We obtain these results by conditioning the photon registration by APD2 on the detection of two photons by APD1. We observe a more pronounced departure from the negative exponential distribution than that observed for one-photon-subtracted thermal light. The mean occupation number increases from $1.14$ without conditioning to $2.54$, and SNR increases from $0.86$ to $1.45$. In parts b and c we plot the mean and SNR as functions of  $\varphi$. Besides the excellent agreement between the theory and experiment, these results confirm the increase in enhancement due to two-photon subtraction. We note that subtraction of a larger number of photons should lead to a further increase in both average occupation number and SNR. 

 \begin{figure}[htbp]
\includegraphics[width=0.85\columnwidth]{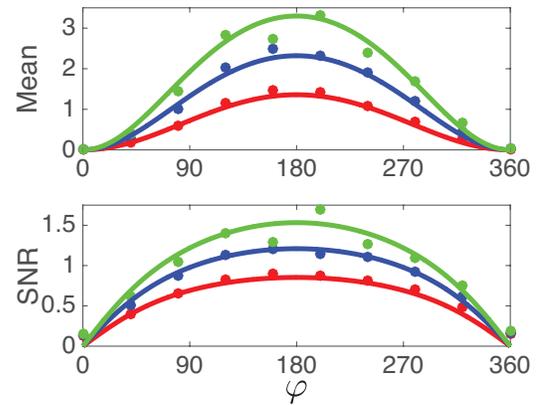} \vspace{-0.3cm}
\caption{The average occupation number (top) and the signal-to-noise ratio (bottom) at the other output port $\hat{d}$ of the MZI measured as a function of the phase shift introduced within the interferometer. The dots represent the experimental results and the lines are the theoretical predictions. Here $\bar{n} = 4.1$, $T=0.9$, $\eta_3\sim 0.42$, and $\eta_1\sim 0.33$. Also included are the non-conditioned, and one-photon-subtracted thermal light for comparison.
}
\label{fig4}
\end{figure}

The above results demonstrate the increase in the mean photon number and SNR of the light from output port $c$ conditioned on the subtraction photons from the same output port. Significantly, theory predicts that photon subtraction from port $c$ also enhances the mean photon number and SNR for light leaving from port $d$. We show results for this situation in Fig.\ \ref{fig4}. We see that the mean photon number and SNR of port $d$ are increased by the process of photon subtraction from port $c$, and that the increase becomes more pronounced for the subtraction of two photons. This surprising dependence of the photon number distribution involving the two ports is a manifestation of the correlations between the output ports $c,d$ in the joint photonic density matrix.  

Although our results show that the SNR of the detection process can be increased through use of photon subtraction, we note that in most situation this increase does not lead to a decrease in the total time required to perform a sensitive interferometric measurement. The reason is that our method uses the probabilistic subtraction of photons from the incident light field, induced by PBS2 of Fig.\ 1.  The reflectivity of this beam splitter needs to be kept small to avoid the possibility of two photons being removed from the light beam within the deadtime of APD1.  For this reason, a large fraction of the time no photon subtraction occurs, and as a result no information about the quantum light state is obtained once post-selection is implemented. When this loss of information is taken into account, it is likely that there is no net increase in the SNR of the measurement process. There is some hope that enhanced detection protocols can be used to overcome this problem.  One is to use other approaches to photon subtraction that are less probabilistic \cite{Rosenblum:2016jr}. At present it is not clear how much overall improvement can be obtained from such a procedure.

In summary, we have introduced and realized a procedure based on photon-subtraction for increasing the SNR with which one can measure the phase shift induced on a thermal light field. We have implemented this method for a single-transverse-mode light field.  This procedure could be generalized to increase the measurement sensitivity for each spatial mode of a multimode light field, a procedure that holds great promise for increasing the sensitivity of image formation of objects illuminated only by weak thermal light fields.

\begin{acknowledgements}
We acknowledge funding from the Defense Advanced Research Projects Agency (DARPA). The views, opinions, and/or findings contained in this presentation are those of the authors and should not be interpreted as representing the official views or policies of the Department of Defense or the U. S. Government. 
\end{acknowledgements}

\bibliographystyle{apsrev4-1}
\bibliography{boeing_refs}

\clearpage

\section{Supplementary materials}

\subsection{Photon subtraction}
\begin{figure}[htbp]
\includegraphics[width=\columnwidth]{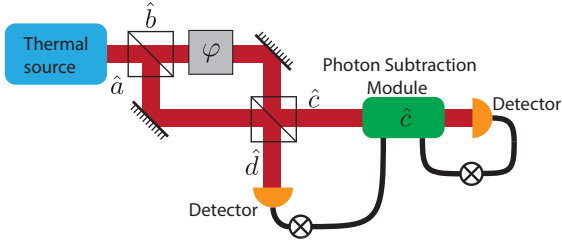}
\caption{Conceptual representation of a thermal interferometry experiment to observe the enhancement of signal and signal-to-noise ratio as result of photon subtraction. }
\label{sup_fig}
\end{figure}

In the following we present a detailed calculation of the effect of the subtraction on the quantum state of a photon that leaves a Mach-Zehnder interferometer.  The two input ports of the interferometer ($a,b$) are fed by a thermal state and the vacuum state respectively (See Fig.\,\ref{sup_fig}). Let us first derive the the reduced density matrix at the two output ports ($c,d$). A schematic version of the model is given in Fig.\,\ref{sup_fig}
The output ports are related to the input ports through a unitary transformation:
\begin{align}\nn
\hat{c} &=\frac{1}{2} \left\{iy \hat{a}-x\hat{b} \right\}\\
\hat{d} &=\frac{1}{2} \left\{i y \hat{b}  + x \hat{a} \right\}
\end{align}
where $x = 1-e^{i\varphi}$, and $y = e^{i\varphi}+1$. The input state is given by 
\begin{align}
\hat{\rho}_0 = \sum_n  \frac{\nbar^n}{(1+\nbar)^{n+1}}\ket{n}_a~_a\bra{n}\otimes \ket{0}_b~_b\bra{0}.
\end{align}
The reduced density matrix of the state at port $c$ can be derived by taking a partial trace on $d$:
\begin{align}
\hat{\rho}_{0c} = {\rm Tr}_d[\hat{\rho}_0 ] = \sum_{m=0}^{\infty} ~_d\bra{m} \hat{\rho}_0  \ket{m}_d
\end{align}
To evaluate this density matrix we need to calculate the following quantity
\begin{align}\nn 
\mathcal{A}_{m,n}&=~_d\bra{m}\otimes ~_c\bra{n} \hat{\rho}_0 \ket{n}_c \otimes \ket{m}_d  \\
&= \frac{1}{m! n!} ~_d\bra{0}\otimes ~_c\bra{0} d^m c^n \hat{\rho}_0  c^{\dagger n}d^{\dagger m}\ket{0}_c \otimes \ket{0}_d 
\end{align}
Note that $\ket{0}_c \otimes \ket{0}_d = \ket{0}_a \otimes \ket{0}_b$, and we can expand this expression to evaluate it; 
\begin{align}\nn 
\mathcal{A}_{m,n}&= \frac{1}{m! n! 4^{m+n}} ~_a\bra{0}\otimes ~_b\bra{0} (-x\hat{b}+iy\hat{a})^n \\ \nn
 &~~~~~(iy\hat{b}+x\hat{a})^m \hat{\rho}_{th}^{(a)}\otimes \ket{0}_b ~_b\bra{0} 
 (-iy\hat{b}^{\dagger}+x\hat{a}^{\dagger})^{m} \\ \nn
 &~~~~~(-x\hat{b}^{\dagger}-iy\hat{a}^{\dagger})^{ n}\ket{0}_a \otimes \ket{0}_b \\ \nn
 &=\frac{|x|^{2m} |y|^{2n}}{m! n! 4^{m+n}} ~_a\bra{0} \hat{a}^{n+m} \hat{\rho}_{th}^{(a)} \hat{a}^{\dagger(n+m)}
 \ket{0}_a\\ \nn
 &=\frac{|x|^{2n} |y|^{2m}}{m! n! 4^{m+n}} \sum_i \frac{\nbar^i  |~_a\bra{0} \hat{a}^{n+m} \ket{i}_a|^2}{(1+\nbar)^{i+1}} \\
 &= \frac{|x|^{2n} |y|^{2m}(m+n)!}{m! n! 4^{m+n}} \frac{\nbar^{m+n}}{(1+\nbar)^{m+n+1}}
\end{align}
Using this result one can calculate the diagonal elements of the reduced density matrix. 
\begin{align} \nn 
~_c\bra{n}\hat{\rho}_c \ket{n}_c&= \sum_m \mathcal{A}_{m,n} \\ \nn &= \frac{|y|^{2n}\nbar^{n}}{4^n(1+\nbar)^{n+1}}  \sum_m \left(_m^{n+m}\right) \left(  \frac{\nbar |x|^2}{4(1+\nbar)} \right)^m\\ \nn
&=\frac{|y|^{2n}\nbar^{n}}{4^n(1+\nbar)^{1+n}}   (\frac{1}{1- \frac{\nbar |x|^2}{4(1+\nbar)}})^{n+1}\\
&=\frac{|y|^{2n}\nbar^{n}}{4^n}   \frac{1}{(1+ \frac{\nbar |y|^2}{4})^{n+1}}
\end{align}
Furthermore one can readily confirm that $\bra{i}\hat{\rho}_c\ket{j}$ vanishes if $i\neq j$. Thus the density matrix at port $\hat{c}$ can be written as 
\begin{align}
\hat{\rho}_c = \sum_n \frac{(\bar{n}\cos^2 \frac{\varphi}{2})^n}{(1+\bar{n}\cos^2 \frac{\varphi}{2})^{n+1}} \ket{n}_c ~_c\bra{n}
\end{align} 
which a thermal state with the reduced occupation number and standard deviation of 
\begin{align}
\bar{n}_c &= {\rm Tr}[\hat{c}^{\dagger}\hat{c} \hat{\rho}_c] =
\bar{n}\cos^2 \frac{\varphi}{2},~~~~~
&\sigma_c= \sqrt{\bar{n}_c^2+\bar{n}_c}
\end{align}
Similarly one can show that the reduced density matrix at port $\hat{d}$ is a thermal state with the reduced occupation number and standard deviation of 
\begin{align}
\bar{n}_d &= {\rm Tr}[\hat{d}^{\dagger}\hat{d} \hat{\rho}_0] =\bar{n}\sin^2 \frac{\varphi}{2},~~~~~
&\sigma_d= \sqrt{\bar{n}_d^2+\bar{n}_d}
\end{align}

Next we study the effect of photon subtraction on the reduced density matrices at output ports. Subtracting a photon in port $\hat{c}$ can be described by the following operation:
\begin{align}
\hat{\rho}_0\rightarrow \hat{\rho}_1 = \frac{\hat{c}\hat{\rho}_0 \hat{c}^\dagger}{\text{Tr}[\hat{c}\hat{\rho}_0 \hat{c}^\dagger]} .
\end{align}
By taking partial trace one can then find the reduced density matrix at each of the output ports.
\begin{align} \nn
\hat{\rho}_{1c} = {\rm Tr}_d [\hat{\rho}_1] .
\end{align}
First we note that 
\begin{align}
~_d\bra{m}\otimes ~_c\bra{n} \hat{\rho}_1 \ket{n}_c \otimes \ket{m}_d = \frac{n+1}{\bar{n}_c} \mathcal{A}_{m,n}
\end{align}
Thus the diagonal elements of the reduced density matrix can be found as following
\begin{align}
~_c\bra{n}\hat{\rho}_c \ket{n}_c&=  \frac{n+1}{\bar{n}_c} \sum_m \mathcal{A}_{m,n}= \frac{(n+1)\nbar_c^{n+1}}{(1+\nbar)^{n+2}}
\end{align}
One can readily confirm that the off-diagonal elements are all zero and thus the reduced density matrix in port $\hat{c}$ after the subtraction are given by
\begin{align}
\hat{\rho}_{1c} = \sum_n  \frac{(n+1)\nbar_c^{n+1}}{(1+\nbar)^{n+2}}  \ket{n}_c ~_c\bra{n}.
\end{align}
The average occupation number and the standard deviation of this distribution are 
\begin{align}\nn
&{\rm Tr}[\hat{c}^{\dagger}\hat{c} \hat{\rho}_1] = 2\bar{n}\cos^2 \frac{\varphi}{2}=2\nbar_c\\
 &({\rm Tr}[(\hat{c}^{\dagger}\hat{c})^2 \hat{\rho}_1]- {\rm Tr}[\hat{c}^{\dagger}\hat{c} \hat{\rho}_1]^2)^{1/2} = \sqrt{2} \sigma_c
\end{align}
Note that the standard deviation increase only by a factor of $\sqrt{2}$ where as the signal has increased by a factor of $2$, and as the result the signal-to-noise ratio is enhanced by a factor of $\sqrt{2}$. One can show that the same result holds for the other output port of the interferometer. Similarly one can show that the conditioned on subtracted events in mode $\hat{c}$ the average occupation number in port $\hat{d}$ doubles and the standard deviation is enhanced by a factor of $\sqrt{2}$ too.
\begin{align}\nn
&{\rm Tr}[\hat{d}^{\dagger}\hat{d} \hat{\rho}_1] = 2\bar{n}\sin^2 \frac{\varphi}{2}=2\nbar_d\\
 &({\rm Tr}[(\hat{d}^{\dagger}\hat{d})^2 \hat{\rho}_1]- {\rm Tr}[\hat{d}^{\dagger}\hat{d} \hat{\rho}_1]^2)^{1/2} = \sqrt{2} \sigma_d
\end{align}
We emphasize that this surprising result could be expected since, in contrast with the input ports, the density matrix at the output ports are not correlated. 

\subsection{Common path Mach-Zehnder interferometer}
In a conventional Mach-Zehnder interferometer the first beam splitter separates the beam into to parts. Each part takes a separate path and and then we bring the two paths together and recombine them using another beam splitter. The difference between the accumulated phase of the paths determines the intensity distribution at the two output ports. The challenging aspect of an MZI is the stability; A slight instability in any of the components would lead to a phase instability and diminishes the fringe visibility. To alleviate this problem one can replace the two spatially separated paths by polarization use wave plates to induce a phase between the two polarizations. As such one no longer needs to spatially separate the two polarizations and considerably mitigate stability of the system. 

\begin{figure}[t!]
\includegraphics[width=8cm]{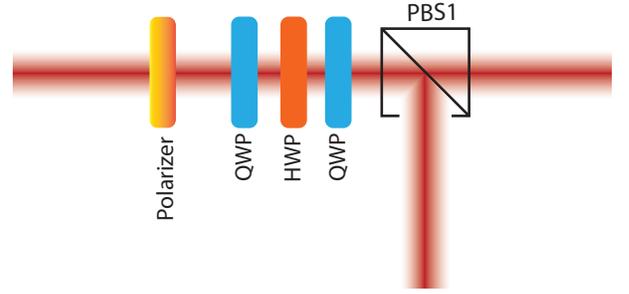}
\caption{A schematic representation of a common-path Mach-Zehnder interferometer that induces a variable phase between the two polarizations.}
\label{CMZI}
\end{figure}

Below we present a detailed discussion on how this interferometer works. In Fig.\,\ref{CMZI} we present a schematic representation of a common-path Mach-Zehnder interferometer. The polarizer prepares the polarization state $\ket{H}$ which can be written as an equally wighted coherent superposition of $\ket{D}$ and $\ket{A}$. Note that here 
\begin{align}\nn 
\ket{D} &=\frac{\ket{H}+\ket{V}}{\sqrt{2}},\\
\ket{A} &=\frac{\ket{H}-\ket{V}}{\sqrt{2}}.
\end{align}
Our aim is to induce a phase between the two components $\ket{D}$ and $\ket{A}$. Then we set a quarter wave-plate (QWP) in a $45^\circ$ angle. The QWP maps $\ket{D}\rightarrow \ket{R}$, and  $\ket{A}\rightarrow \ket{L}$ where
\begin{align}\nn
\ket{R} &=\frac{\ket{H}+i\ket{V}}{\sqrt{2}},\\
\ket{L} &=\frac{\ket{H}-i\ket{V}}{\sqrt{2}}.
\end{align}
Next we can use a half wave-plate that induces a variable phase between the two components $\ket{R}$ and $\ket{L}$. Again a QWP can be used to map  $\ket{R}\rightarrow \ket{D}$ and $\ket{L}\rightarrow \ket{A}$, and finally we use a polarizing beam splitter to separate the two polarizations $\ket{D}$ and $\ket{A}$. By rotating the HWP we can change the induced phase between the two polarizations $\ket{D}$ and $\ket{A}$ and we get a one-to-one mapping between this setup and a conventional Mach-Zehnder interferometer. 
\subsection{Surjective photon counting}
We emphasize that the effect of the inherent quantum efficiency of the detector can be modeled by combination of a beam splitter and a detector of quantum efficiency of $100\%$. That is a detector that fires if at least one photon arrives. Thus for simplifying our analysis we assume a detector with detection efficiency of $100\%$. For thermal light with average occupation number of $\bar{n}$ the probability of incurring an $N$-photon event is given by 
\begin{align}
P(N) = \frac{\nbar^N}{(1+\nbar)^{N+1}}.
\end{align}
In our surjective detection scheme this event may be registered as a detection of a lower number of photons if more than one photons arrive separated by less than the dead time of the detector. Assuming that the dead time of the detector time is $\sim 50\,ns$ and the coherence time of the source is $\sim\,1\mu s$ in each coherence time there are $K = 20$ time bins. In principle an $N$-photon event can be registered as any of $\{1,2,\cdots,N\}$-photon events, and since we work with very few photons we assume that always $K>N$. Then the probability distribution of number of clicks if $N$ photons arrive in a temporal mode can be cast as a combinatorics problem and one uses Bayes' theorem:
\begin{align}
P(m) = \sum_N P(N)P(m|N).
\end{align}
\begin{figure}[htbp]
\includegraphics[width=8cm]{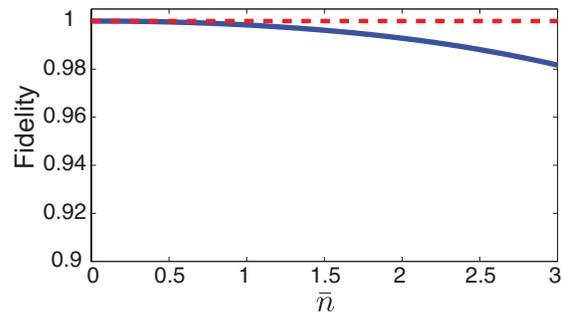}
\caption{The fidelity between the probability distribution of thermal statistics and the probability distribution detected by our photon counting scheme for thermal lights of different values of average occupation number.}
\label{fig4}
\end{figure}
to find the modified probability distribution, $P(m)$, that the APDs register. In Fig.\,\ref{fig4} we plot the fidelity between the probability distribution of the thermal statistics and the probability distribution detected by the our surjective counting scheme. Fidelity is a measure of distance between any two probability distributions. The fidelity of two probability distributions $\{q_i\}$ and $\{p_i\}$ is defined by $\sum_i \sqrt{p_i q_i}$ and its range between $\{0,1\}$ \cite{Nielsen:2009ga}. The high value of the fidelity confirms our initial intuition that for low photon number, our counting scheme provides an excellent approximation to the actual photon distribution. Finally it should be noted that to compare with the experimental results one can feed the probability distributions that are predicted by the theory into an algorithm that counts for the surjective nature of the counting mechanism before comparing them to the experimental results.

\end{document}